\documentclass[twocolumn,showpacs,preprintnumbers,amsmath,amssymb]{revtex4}
\usepackage{graphicx}
\usepackage{dcolumn}
\usepackage{bm}

\usepackage{epsfig}
\newcommand{\beq}{\begin{equation}}
\newcommand{\eeq}{\vspace{0cm} \end{equation}}
\newcommand{\beqq}{\setlength\arraycolsep{2pt}\begin{eqnarray}}
\newcommand{\eeqq}{\vspace{0cm} \end{eqnarray}}

\begin{document}

\title{Gravitationally Induced Particle Production: Thermodynamics and Kinetic Theory}
\author{J. A. S. Lima$^{1}$\footnote{limajas@astro.iag.usp.br}}
\author{I. Baranov$^{2}$\footnote{iuribaranov@usp.br}}

\affiliation{$^{1}$Departamento de Astronomia, Universidade de S\~{a}o
Paulo \\ Rua do Mat\~ao, 1226 - 05508-900, S\~ao Paulo, SP, Brazil}

\affiliation{$^{2}$Departamento de F\'isica Geral, Universidade de S\~{a}o
Paulo \\ Rua do Mat\~ao, 187 - 05508-090, S\~ao Paulo, SP, Brazil}


\bigskip
\begin{abstract}
A relativistic kinetic description for the irreversible thermodynamic process of gravitationally induced particle production is proposed in the context of an expanding Friedmann-Robertson-Walker (FRW) geometry. We show that the covariant thermodynamic treatment referred to as ``adiabatic'' particle production provoked by the cosmic time-varying gravitational field has a consistent kinetic counterpart. The variation of the distribution function is associated to a non-collisional kinetic term of quantum-gravitational origin which is proportional to the ratio $\Gamma/H$, where $\Gamma$ is the gravitational particle production rate and H is the Hubble parameter. For $\Gamma << H$ the process is negligible and as should be expected it also vanishes (regardless of the value of $\Gamma$) in the absence of gravitation. The resulting non-equilibrium distribution function has the same  functional form of equilibrium with the evolution laws corrected by the particle production process. The macroscopic temperature evolution law is also kinetically  derived  for massive and massless particles. The  present approach points to the possibility of an exact (semi-classical) quantum-gravitational  kinetic treatment by incorporating back-reaction effects in the cosmic background.  
\end{abstract}

\maketitle

\section{Introduction}

Many decades ago, Parker and collaborators put forward a microscopic  description to the particle production mechanism due to the time varying gravitational field of the expanding Universe \cite{Parker}. Such studies were based in the so-called  Bogoliubov mode-mixing
technique in the context of quantum field theory (QFT) in curved spacetime. 

Qualitatively, this gravitationally induced quantum creation mechanism  can easily be understood. In the case of a scalar field evolving in a FRW geometry, for example, its effective mass becomes a time dependent quantity. When the field is quantized, this leads to particle creation with the energy for the newly created particles being supplied by the time-varying gravitational background \cite{BirrellD}. In other words, unlike the Minkowski spacetime, the time varying geometry behaves like a `pump' transforming curvature into particles. 

Particle production in the expanding Universe has also been intensively investigated in the context of inflation,  in particular, during the reheating  and preheating stages.  The former is an extreme  nonadiabatic process happening just after the (adiabatic) slow-rolling regime while the latter is characterized by a highly nonthermal exponential particle production due to a parametric resonance in the first stages of the reheating process \cite{reheating,reheating1}. Indeed, particle production may also take place even during a possible non-adiabatic slow-rollover phase as assumed in the so-called warm inflationary scenarios \cite{Berera}.

On the other hand, although describing rigorously the particle production process for test fields (scalar, vectorial or tensorial) evolving in the cosmic background this approach does not provide a clear recipe of how the created particles modify the classical Einstein field equations (EFE). This is the famous quantum back-reaction problem, a subject that although intensively investigated in the literature has not been properly solved yet. 

Later on, the emergence of particles in the spacetime at the expenses of the gravitational field  was also macroscopically  described by Prigogine and coworkers based on the non-equilibrium  thermodynamics of open systems \cite{Prigogine}. Such an approach was rediscussed by some authors \cite{LCW} within a manifestly covariant formulation and further applied to cosmology. Its basic advantage is that back-reaction is naturally incorporated and constrained by the second law of thermodynamics.  

Nevertheless, it is also an incomplete description in the sense that the matter creation rate must be calculated from QFT in the FRW geometry. The difference between gravitationally induced particle production  and the  bulk viscosity mechanism (as phenomenologically suggested by Zeldovich \cite{ZED} to describe particle creation) was also discussed by Lima and Germano \cite{LG92}. These authors demonstrated that both scalar dissipative processes (particle production and bulk viscosity) are able to generate the same cosmic dynamics but are fully different from a thermodynamic point of view. 

In the last few years, much attention has been paid for cosmologies driven by gravitational  ``adiabatic" particle production where matter and entropy are generated but the specific entropy  (per particle) remains  constant \cite{Models,LSS08,LJO2010,Pert11,LBC12,MP13,Komatsu2014,Waga2014,LGPB}. Since the creation process is macroscopically described by a negative pressure of quantum origin (back reaction), these models have been investigated  as a possibility to unify the so-called cosmological dark sector.  In this case,  as suggested long ago, the  accelerating stage of the Universe at all phases (early and late time inflation)  becomes a consequence of the induced  gravitational  particle production \cite{Models}.  The connection between warm inflation \cite{Berera}, decaying $\Lambda$ models and particle production cosmologies has been discussed by several authors\cite{PLB2014,Others}.

Along these lines, it should be recalled that Lima et al. \cite{LJO2010} proposed a creation cold dark matter (CCDM) cosmology with one free parameter that is equivalent to the $\Lambda$CDM evolution both at the background and perturbative levels \cite{Waga2014} (see, however, \cite{Pert11}).  More recently, even a complete cosmology where the space-time-matter evolves between an early and a late time de Sitter phases driven by particle production has been proposed and its predictions compared with the available astronomical data \cite{LBC12}. The  consistence of such a scenario with the generalized second law of thermodynamics was also discussed in detail  by Mimoso and Pav\'on \cite{MP13} (see also \cite{LGPB} for a more general analysis).  

Indeed, although very successful to mimic $\Lambda$CDM model predictions, it has also been recognized since long ago that the lack of a more fundamental approach for this macroscopic treatment works like a Damocles sword hanging over the foundations of any cosmology endowed with 
continuous gravitationally induced particle creation.

In this paper,  we go one step further by examining the kinetic basis of the macroscopic description for particle  production at the expenses of the gravitational field in the framework of the FRW geometry.  In order to clarify some subtleties present in the earlier results, we first rediscuss the irreversible macroscopic thermodynamic description in the case of ``adiabatic" particle production. 

As we shall see,  associated with such thermodynamics there is a consistent microscopic kinetic description from which all macroscopic nonequilibrium results are readily recovered. Unless explicitly stated, in this work we use units in which the speed of light and the Boltzmann constant are $c=k_{B}=1$.  


\section{Cosmology, Thermodynamics and Matter Creation}

In what follows we consider that the Universe is described by an arbitrary FRW geometry

\begin{equation}
 ds^2 = dt^2 - a^{2}(t)\left(\frac{dr^2}{1-kr^{2}} +  r^{2}d\theta^2  +  r^{2}sen^2\theta d\phi^{2}\right),
\end{equation}
where $a(t)$ is the scale factor and $k=0, \pm 1$ is the curvature parameter. For the sake of completeness as well as to specify the kind of creation process 
discussed here we first review briefly  the thermodynamic description in such a  background.

\subsection{Equilibrium states}

Following standard lines, the equilibrium thermodynamic states of a
relativistic simple fluid are characterized by three independent quantities: a particle current $N^{\mu }$, an
entropy current $S^{\mu}$, and an energy momentum
tensor $T^{\mu \nu}$ which satisfy  the following  conservation laws:
\begin{equation} \label{eq:NA}
N^{\mu}=nu^{\mu}, \quad  N^{\mu};_{\mu}=0,
\end{equation}
\begin{equation} \label{eq:SA}
S^{\mu}=s u^{\mu}, \quad  S^{\mu};_{\mu}=0,
\end{equation}
\begin{equation}\label{eq:TAB}
T^{\mu \nu}=(\rho + P)u^{\mu} u^{\nu} - Pg^{\mu \nu},
\quad T^{\mu \nu};_{\nu}=0, 
\end{equation}
where ($;$) means covariant derivative, $\rho$, $P$, $n$ and $s$, are the energy density, pressure, particle number
density, and entropy density, respectively. In the FRW background, the
above conservation laws can be rewritten as (a dot means comoving
time derivative)

\begin{equation}\label{CL}
 \dot{\rho} + (\rho + P)\Theta=0,\,\,
 \dot{n} + n\Theta=0,\,\, \dot{s} + s\Theta=0,
\end{equation}
where the scalar of expansion, $\Theta=3\dot a/a\equiv 3H$ (H is the Hubble parameter). The quantity $\Theta^{-1}$ specify the macroscopic time scale of the fluid. 

The above thermodynamic  quantities [$\rho$, P, n and s] are related to the
temperature $T$ by the local Gibbs law:

\begin{equation} \label{eq:GIBBS}
nTd\left( \frac{s}{n} \right)\equiv nTd\sigma = d\rho - {\rho + P \over n}dn,
\end{equation}
where $\sigma$ is the entropy per particle. In addition, the validity of the local form of Euler
relation \cite{callen} 
\beq 
nT\sigma={P +\rho}-{\mu n},\,\label{entropyd}
\eeq
where  $\mu$ is the chemical potential together the Gibbs law leads to the so-called Gibbs-Duhem relation: 
\beq 
n\sigma dT= dP - nd\mu, \label{GB}
\eeq
showing that there are only two
independent thermodynamic variables, say, $n$ and $T$.
Now, by assuming that $\rho=\rho(T,n)$ and $P=P(T,n)$  and combining the energy conservation laws with  the
thermodynamic identity 

\begin{equation}
T \biggl({\partial P \over \partial T}\biggr)_{n}=\rho + P - n
\biggl({\partial \rho \over \partial n}\biggr)_{T},
\end{equation}
one obtains the temperature law 
\begin{equation} \label{eq:EVOLT0}
{\dot T \over T} = \biggl({\partial P \over \partial
\rho}\biggr)_{n} {\dot n \over n}\equiv -\biggl({\partial P \over \partial
\rho}\biggr)_{n}\Theta.
\end{equation}

It should  be stressed that the above temperature
evolution law presented here is fully independent of the entropy
function, as well as, of the chemical potential $\mu$. In particular,  for radiation or ultrarelativistic particles of mass m ($m << T$, $\rho \simeq 3p$), the above temperature evolution law becomes
\begin{equation} \label{eq:EVOLT}
{\dot T \over T} = -{\dot a \over a}.
\end{equation}
A direct integration of this expression furnishes $aT=a_0T_0$, or equivalently $T(z)=T_0(1+z)$ (temperature Friedmann's prediction), where $z = a_0/a - 1$ is the redshift parameter. As usual, $a_0$ and $T_0$  are, respectively,  the  present day values of the scale factor and temperature of cosmic microwave background radiation (CMBR). Let us now discuss how this basic formalism is modified in the presence of gravitationally induced particle production.

\subsection{Non-equilibrium states due to particle creation}\label{2b}

Let us now assume that particles are spring up into the homogeneous and isotropic expanding Universe  due to the cosmic time-varying gravitational field. 

To begin with we recall that the emergence of particles means that all the balance equilibrium equations (\ref{eq:NA})-(\ref{eq:TAB}) must be modified. In other words, the divergence of the basic equilibrium quantities like the particle and entropy fluxes and energy-momentum tensor  are now different from zero. However, from EFE  ($G^{\mu \nu}=8\pi G T^{\mu \nu}$), we know that the total energy-momentum tensor (EMT) must be conserved. As happens in the case of the classical bulk viscosity mechanism (or more generally with all transport processes), this problem is solved by assuming that any non-equilibrium correction is incorporated in the complete EMT which is identically conserved. 

In the case of a scalar process here represented by a uniform distribution of emerging particles the new contribution for the EMT is formed by a dynamic non-equilibrium  pressure. However, the definition of the particle and entropy fluxes are not modified even considering that such quantities are not conserved.   

In the presence of a gravitational particle source, the balance equation for the particle flux becomes (note that in the old notation of Refs. \cite{LCW,LG92}, $n\Gamma\equiv\Psi$).

\begin{equation}\label{particleflux}
 N^\mu;_{\mu} \equiv \dot{n} + n\Theta=n\Gamma,
\end{equation}
where $\Gamma$ is the particle production rate ($Dim\,[\Gamma]\equiv time^{-1}$). Defining the total number of particles in the comoving volume by, $N=n a^{3}$, the above balance equation can be rewritten as:
\begin{equation}\label{NEq}
\frac{\dot N}{N} = \Gamma,
\end{equation}
showing again that $\Gamma$ drives the particle creation rate in the comoving volume. Obviously, when compared with $\Theta$ this new microscopic time scale quantifies  the efficiency of the gravitational particle production. In particular, if $\Gamma << \Theta$ the creation process can safely be  neglected. 

The entropy flux vector is defined by
\begin{equation}
S^\mu=su^\mu\equiv n\sigma u^\mu,
\end{equation}
and by taking its divergence one finds
\begin{equation}
 S^\mu;_{\mu}=n \dot \sigma + \sigma n \Gamma,
\end{equation}
Hence, if the creation process occurs in such a way that the specific 
entropy is constant (``adiabatic"' gravitational particle creation), the divergence of $S^{\mu}$ reduces to (see Refs. \cite{LCW,LG92} for details)
\begin{equation}\label{DivS}
 S^\mu;_{\mu}= n\sigma\Gamma \equiv s\Gamma.
\end{equation}
Note also that in the homogeneous and isotropic case considered here, $\sigma = S/N$, where $S=sa^{3}$ is the entropy in a comoving volume. This means that the condition  $\dot \sigma=0$ has a direct physical meaning, namely:
\begin{equation}
\dot S/S=\dot N/N  \Rightarrow S = k_B N.
\end{equation}
Therefore, unlike the bulk viscosity mechanism, the entropy growth in the gravitational particle production process is closely related with the emergence of particles in the space-time thereby leading to the expected enlargement of the  phase space. Here we are particularly interested in the kinetic description of this non-equilibrium process.

On the other hand, the energy-momentum tensor for a fluid endowed with particle production must also be corrected according to the thermodynamic second law. In principle we may write: 
\begin{equation}\label{correction}
T^{\mu \nu}=  T^{\mu \nu}_{E} + \Delta T^{\mu \nu},
\end{equation}
where $T^{\mu \nu}_{E}$ describes the equilibrium states (see Eq.(\ref{eq:TAB})) and  $\Delta T^{\mu \nu}$ is a correction describing the effects of particle production.
Invoking the isotropy and homogeneity of space such a correction must be represented by a scalar process. In terms of components the possible corrections have the following forms:
\begin{equation}\label{Comp}
\Delta T^{0}_0=0 \,\, and \,\, \Delta T^{i}_j=-P_c\delta^{i}_{j},
\end{equation}
where the first condition is the constraint removing the ambiguity on the energy density for non-equilibrium states ($\rho$ is the same function in the absence of ``dissipation"), and $P_c$ is the dynamic (creation) pressure which describes macroscopically the emergence of particles  in the space-time. In a manifestly covariant description we can write 

\begin{equation}\label{deltaT}
 \Delta T^{\mu \nu} =  -P_c (g^{\mu \nu} - u^\mu u^\nu)\equiv -P_ch^{\mu\nu}, 
\end{equation}
where $h^{\mu \nu}$ is the projector onto the rest frame of $u^\mu$. Naturally, $\Delta T^{\mu \nu}$ works like a source term for the equilibrium EMT which can be incorporated to describe the process through a conserved EMT as required by the Einstein field equations. Actually, the divergence of the above correction reads:
\begin{equation}
u_{\mu}\Delta T^{\mu\nu}_{;\nu}= P_c \Theta,
\end{equation}
whereas the energy conservation law for the complete EMT 

\begin{equation}\label{eq:TABC}
T^{\mu \nu}=(\rho + P + P_c)u^{\mu} u^{\nu} - (P + P_c) g^{\mu \nu},
\end{equation}
becomes
\begin{equation}\label{energy}
u_{\mu}T^{\mu\nu}_{;\nu} = \dot \rho + \Theta(\rho + P + P_c)=0,
\end{equation}
so that if $P_c$ is zero (no particle production), the equilibrium energy conservation law [the first of Eqs. (\ref{CL})] is recovered.

{\it What about the form of $P_c$?} From Eq. (\ref{DivS}) we know that the positivity of the entropy production is already obeyed. On other hand, the local equilibrium principle implies that the local Gibbs relation (\ref{eq:GIBBS}) remains valid even for processes out of equilibrium. Differentiating it under the guise of $\dot \sigma=0$ we obtain with the help of (\ref{particleflux}) 
\begin{equation}\label{dotrho}
\frac{\dot \rho}{\rho + P}=\frac{\dot n}{n} = \Gamma-\Theta,
\end{equation}
and inserting the above relation into (\ref{energy}), one gets the ``adiabatic' creation pressure formula (see \cite{LCW} for a more general deduction including $\dot \sigma \neq 0$)

\begin{equation}\label{pressaotermo}
 P_c=-(\rho + P)\frac{\Gamma}{\Theta}.
\end{equation}
An important point to keep in mind here is that the presence of the creation pressure in this macroscopic description is not the result of  collisional processes as happens, for instance,  with the bulk viscosity mechanism. 

{\it How the temperature evolution law is modified by the creation process?}  It is easy to show that the temperature evolution for a system with matter creation reads \cite{LCW,LG92} 
\begin{equation}\label{TEMP}
\frac{\dot T}{T}=-\left(\frac{\partial P }{\partial \rho} \right)_n \Theta-\frac{P_c\Theta + n\Gamma(\partial \rho / \partial n)_T}{T(\partial \rho / \partial T)_n}.
\end{equation}
Therefore, if $\dot \sigma = 0$, we may insert (\ref{pressaotermo}) in the above equation, leading to
\begin{equation}\label{Templaw}
 \frac{\dot T}{T}=\left( \frac{\partial P}{\partial \rho}\right)_n\frac{\dot n}{n}=\left( \frac{\partial P}{\partial \rho}\right)_n\left(\Gamma - \Theta\right),
\end{equation}
which reduces to the equilibrium law in the limit $\Gamma \rightarrow 0$ (see Eq. (\ref{eq:EVOLT0})). Now,  for ultrarelativistic particles or radiation, the above temperature evolution law yields 
\begin{equation}\label{temp2}
 \frac{\dot T}{T}=-\frac{\dot a}{a}+\frac{\Gamma}{3}, 
\end{equation}
which clearly departs from Friedmann's law ($T \propto a^{-1}$) for $\Gamma \neq 0$ (cf. Eq. (\ref{eq:EVOLT})).  

It is widely believed that the cosmological creation of photons in the expanding Universe is not allowed  because the blackbody form of the spectrum is destroyed \cite{SW}. However, as discussed long ago by one of the authors \cite{Lima97}, the blackbody form of the CMB spectrum is also preserved when the production of photons happens under ``adiabatic" conditions.  On other words, the particles spring up into space-time with the same temperature of the already existing ones as a consequence of the condition $\dot \sigma = 0$.  Physically, this happens because the form of equilibrium temperature law as a function of the thermodynamic variables is preserved during the expansion (see the first equality in Eq. (\ref{Templaw})) and discussion below Eq. (\ref{energyL})).

More recently, using a phenomenological expression for $\Gamma$ \cite{Lima97,LGA96}, different authors have discussed some consequences of the above law based on the Sunyaev-Zeldovich effect and other cosmological observations \cite{TempL}. 

In the nonrelativistic limit, $\rho  \simeq nm + 3nT/2$ and $P=nT$. In this case  the temperature law (\ref{Templaw}) assumes the form:
\begin{equation}\label{temp3}
\frac{\dot T}{T}=-2\frac{\dot a}{a}+\frac{2}{3}\Gamma, 
\end{equation}
which for $\Gamma=0$ reduces to the standard result, $T \propto a^{-2}.$

The general integration of (\ref{temp2}) and (\ref{temp3}),  as well as, a brief discussion of some specific models will be postponed to section V when the same temperature laws it will be deduced based on the kinetic approach. 

At this point, we would like to stress that a consistent kinetic treatment describing the gravitationally induced particle production process as discussed here must recover Eqs. (\ref{particleflux}), (\ref{DivS}), (\ref{pressaotermo}), (\ref{temp2}) and (\ref{temp3}).  This is the main aim of the next sections. 
\section{Boltzmann's Equation and Gravitational Particle Production}

Let us now discuss the kinetic counterpart for the above irreversible description of gravitationally induced particle production. 
We will start by investigating a new possible form for the extended Boltzmann equation when the referred process is taken into account.  

\subsection{The kinetic non-collisional term}
First we recall that the one particle distribution function $f(x^{\mu}, p^{\mu})$ for non-equilibrium processes  must be described by an extension of the standard relativistic Boltzmann equation \cite{Stewart,Bernstein}. The first attempt to incorporate gravitational particle production was discussed long ago by  Triginer, Zimdahl and Pav\'on \cite{TZP96a}. Following these authors we starting by writing the modified Boltzmann equation in the form: 

\begin{equation}\label{boltzmanneq}
L(f) \equiv p^\mu\frac{\partial f}{\partial x^\mu}-\Gamma^{\mu}_{\alpha \beta}p^\alpha p^\beta \frac{\partial f}{\partial p^\mu}={\mathcal C }(f) + {\mathcal P}_g (x^{\mu},p^{\mu}),
\end{equation}
where $f(x^\mu, p^\mu)$ is the distribution function, $\Gamma^{\mu}_{\alpha \beta}$ are the Cristoffel symbols, and  ${\mathcal C}(f)$ is the standard collisional term which for an expanding uniform gas is responsible for the bulk viscosity mechanism. 

The new source term,  ${\mathcal P}_g$, is an unknown  non-collisional contribution of gravitational-quantum origin. It encodes the gravitationally induced creation process associated with the time varying gravitational field of the cosmic expanding Universe.  

In what follows we fully neglect ${\mathcal C}(f)$ for two distinct reasons: (i) its consequences have already been extensively investigated in the literature, and, more important, (ii) Unlike the standard collisional contributions, the new ${\mathcal P}_g$ term has an unknown quantum (non-collisional) nature, and, us such, its existence requires an alternative treatment which, in principle, is conceptually different from the standard Boltzmann approach.

It is also worth noticing that the Boltzmann equation (\ref{boltzmanneq}), including collisional and gravitational production terms is a general one in the sense that the mass shell condition  has not been imposed. When dealing within the mass-shell - a physical constraint implying that  $f\equiv f(x^\mu,~p^i)$ - the above Boltzmann's equation (neglecting the standard collisional term) can be written as:
\begin{equation}\label{restricted}
 p^\mu\frac{\partial f}{\partial x^\mu}-\Gamma^{i}_{~\alpha \beta}p^\alpha p^\beta \frac{\partial f}{\partial p^i}={\mathcal P}_g,
\end{equation} 
where the possible contributions to $\mathcal{P}_g$ are also restricted to the mass-shell (the Latin index take values $i=$1,2,3).

Before to proceed further, it is useful to rewrite the Boltzmann's equation in terms of the physical momentum. From now on, in order to simplify matters we assume a flat FRW geometry 
\begin{equation}
 ds^2 = dt^2 - a^{2}(t)\left(dx^2 +  dy^2  +  dz^2\right),
\end{equation}
for which the non-null Cristoffel symbols are \cite{SW}

\begin{equation}\label{symbols}
\Gamma^{0}_{00}=\Gamma^{i}_{jk}=0, \,\,\,\Gamma^{i}_{0j}= \frac{\dot a}{a}\delta^{i}_{j},\,\,\, \Gamma^{0}_{ij} = -\frac{\dot a}{a}{g_{ij}}. \,\,\,
\end{equation}

The physical momentum ($\bar p^{i}$) is related to the comoving momentum ($p^{i}$) by $\bar{p}^i = ap^i$. The Boltzmann's equation in terms of physical momentum and coordinates reads:
\begin{equation}\label{PM}
 \frac{df(\bar{x}^\mu,~\bar{p}^i)}{d\lambda}=\bar{p}^\mu\frac{\partial f}{\partial \bar{x}^\mu}+\frac{\partial f}{\partial \bar{p}^i}\frac{d\bar{p}^i}{d\lambda}={\mathcal P}_g  .
\end{equation}
Where $\frac{d\bar{p}^i}{d\lambda}$ is related with the geodesic equation by
\begin{equation}
 \frac{d\bar{p}^i}{d\lambda}=a \frac{dp^i}{d\lambda}+p^i\frac{da}{d\lambda}=a \frac{dp^i}{d\lambda}+\dot a p^ip^0.
\end{equation}
The comoving momenta obey the geodesic equation
\begin{equation}
 \frac{dp^i}{d\lambda}=-\Gamma^i_{~\alpha \beta}p^\alpha p^\beta,
\end{equation}
and inserting these results into (\ref{PM}), we find

\begin{equation}
 \bar{p}^\mu\frac{\partial f}{\partial \bar{x}^\mu}-\left(a\Gamma^i_{~\alpha \beta}p^\alpha p^\beta-\dot a p^ip^0\right)\frac{\partial f}{\partial \bar{p}^i}={\mathcal P}_g.
\end{equation}

In this case, the Boltzmann's equation  (\ref{restricted}) assume the final form adopted here:
\begin{equation}
 \bar{p}^0\frac{\partial f}{\partial t}-\frac{\dot a}{a}\bar{p}^0\bar{p}^i\frac{\partial f}{\partial \bar{p}^i}={\mathcal P}_g\,.
\end{equation}

{\it How ${\mathcal P_g}$ can be modeled?} Firstly,  we know that the particles are created from the gravitational field, and, as such,  the new term must take into account gravity. In particular, this means that the net effect must vanish identically in the flat space-time (when the metric tensor takes the global Minkowskian form). Secondly, we know that the intensity of the process depends on the ratio $\Gamma/\Theta$. Therefore, we model ${\mathcal P_g}$ assuming only two conditions: 

(i) $P_g \propto \Gamma^{i}_{\alpha \beta}$  so that the particle production (as described in the mass shell) is forbidden in the absence of gravity 

(ii) ${\mathcal P_g} \propto  \Gamma/\Theta$, a condition naturally  suggested by the macroscopic irreversible approach (see Eqs. (12), (15),(18) and (20)).  

Hence, $P_g$ must be proportional to the product of both quantities, and the simplest ansatz satisfying such requirements are:  

\begin{equation}
{\mathcal P_g}= -\lambda\frac{\Gamma}{\Theta}\Gamma^{i}_{\alpha \beta}\bar{p}^\alpha \bar{ p}^\beta \frac{\partial f}{\partial \bar{p}^i}, 
\end{equation}
where $\lambda$ is a pure number whose specific value can be fixed in order to reproduce exactly the ``adiabatic" macroscopic approach for all corrections appearing in the basic quantities. A  simple algebra for the particle flux shows that $\lambda=1/2$, and, us such, it will be henceforth universally adopted. 

It is worth notice that the above form of the gravitational production creation term has a rather different conception of the one proposed in Ref. \cite{TZP96a}.  There a separation  of ${\mathcal P_g}$ (H in their notation), somewhat inspired in the standard collisional approach was proposed (see their equation (37)).


In what follows we will work only with the physical momenta and for convenience overbars will be dropped. Using the fact that $f$ is isotropic, the extended Boltzmann's equation takes the final form
\begin{equation}\label{boltzmann}
 \frac{\partial f}{\partial t}=H\left(1-\frac{\Gamma}{\Theta} \right)p\frac{\partial f}{\partial p}.
\end{equation}
\section{Adiabatic Particle Production: Kinetic Formulation}\label{kinetic}

In this section we derive the results presented in  section IIb based on the kinetic theory for a comoving observer. For this end, we use the standard definitions of the macroscopic quantities 
\begin{eqnarray}
N^{\mu}&=&\int{fp^\mu\frac{d^3p}{p^0}}, \\ \label{KF1}
S^\mu&=&-\int{(f \ln f - f)p^\mu\frac{d^3p}{p^0}},\label{entropyflux} \\ \label{KF2}
T^{\mu \nu}&=&\int{fp^\mu p^\nu \frac{d^3p}{p^0}}, \label{KF3}
\end{eqnarray}
observing that due to the space isotropy the only nonvanishing components of $N^{\mu}$ and  $S^{\mu}$ are  given  by:
\begin{eqnarray}\label{entropyflux0}
N^{0}&=n=&\int{f{d^3p}},\\
S^{0}&=s=&-\int{(f \ln f - f){d^3p}},
\end{eqnarray}
while the energy momentum has a diagonal form with energy density $\rho=u_{\mu}u_{\nu}T^{\mu\nu} \equiv T^{00}$. This means that any isotropic correction $\Delta T^{\mu\nu}$ to the equilibrium states must have $\Delta T^{00}=0$ (see discussion below Eq. (\ref{correction})).     

Now, with the help of the particle production Boltzmann's operator (\ref{boltzmann}), the divergences of the thermodynamic fluxes will be explicitly calculated.
\begin{itemize}
\item Particle Flux

\begin{eqnarray}
 N^\mu_{; \mu}&\equiv&\frac{1}{a^3}\frac{\partial}{\partial x^\mu}\left( a^3 \int{fp^\mu\frac{d^3p}{p^0}} \right)=\frac{1}{a^3}\frac{\partial}{\partial t}(a^3\int{fd^3p}) \nonumber \\
&=&3Hn+H\left(1-\frac{\Gamma}{\Theta}\right)\int{p\frac{\partial f}{\partial p}d^3p},
\end{eqnarray}
where Eq. (\ref{boltzmann}) was used in the last step. The integral of the last term is easily computed by parts. We find
\begin{eqnarray}
\int{p\frac{\partial f}{\partial p}d^3p}=4\pi \int{p^3\frac{\partial f}{\partial p}dp}=-3\int{fd^3p}=-3n,
\end{eqnarray}
and inserting this result into (34) we obtain the expected result (see Eq. (12))
\begin{equation}
 N^\mu_{;\mu}=n\Gamma.
\end{equation}

\item Entropy Flux
\begin{eqnarray}
S^\mu_{;\mu}&\equiv&-\frac{1}{a^3}\frac{\partial}{\partial x^\mu}\left(a^3 \int{(f \ln f - f)p^\mu \frac{d^3p}{p^0}} \right) \nonumber \\
&=&-\frac{1}{a^3}\frac{\partial }{\partial t}\left(a^3 \int{(f \ln f - f)d^3p} \right), 
\end{eqnarray}
or equivalently, 
\begin{eqnarray}
S^\mu_{;\mu}&=&3sH-\int{\frac{\partial f} {\partial t}\ln f d^3p} \nonumber \\
&=&3sH-H\left( 1-\frac{\Gamma}{\Theta}\right)\int{p\frac{\partial f}{\partial p}\ln f d^3p}.
\end{eqnarray}
The last integral term above can be rewritten in a more convenient way (after a simple integration by parts): 
\begin{eqnarray}
 \int{p\frac{\partial f}{\partial p}\ln f d^3p}&=&4\pi \int{p^3\frac{\partial f}{\partial p}\ln f dp} \nonumber \\
&=&-3\int{(f \ln f - f)d^3p}, 
\end{eqnarray}
and inserting this result into (38) we obtain
\begin{eqnarray}
S^\mu_{;\mu}&=&-\Gamma\int{\left(f\ln f- f \right)d^3p},
\end{eqnarray}
and using again definition (\ref{entropyflux0}) one gets the macroscopic form (see Eq. (\ref{DivS}))
\begin{equation}
S^\mu_{;\mu}=s\Gamma.
\end{equation}
Similarly to what happens  in the macroscopic description of ``adiabatic" particle production (in the sense that ${\dot \sigma}=0$), the above result also suggest that the distribution function of our problem must have the same form of equilibrium, that is, $f(t,p)=e^{\alpha(t) - \beta(t)E}$. Therefore, from now on such distribution function  (the solution of our problem)  it will be termed ``adiabatic" distribution function. In the Appendix A, we provide an alternative deduction of the above result using this ``adiabatic" form.  

\item Energy-Momentum Tensor

Let us now calculate the divergence of the total energy-momentum tensor projected onto $u_\mu$. It involves a rather delicate and subtle aspect from a kinetic viewpoint since the Einstein field equations requires a divergenceless  energy-momentum tensor and we know that the creation pressure has a non-collisional origin. In addition, like the bulk viscosity, we are aware that that the creation pressure  is also negative for an expanding Universe (see Eq. (\ref{pressaotermo}).  However, we recall that  unlike the energy density there is no constraint conditions to the kinetic pressure for states out of equilibrium \cite{Bernstein}. In this way, we can also assume here (like in the macroscopic approach) the existence of a non-collisional corrective pressure term. Now, for the sake of generality, we assume that it takes the following form  $\Delta {\tilde T^{i}_j}  = -{\tilde P}\delta^{i}_j$, or equivalently, ${\Delta {\tilde T^{\mu \nu}}}= -{\tilde P}h^{\mu \nu}$ (${\tilde P}$ is an unknown creation pressure 
which is not necessarily equal to the macroscopic value).  In this case  we  can write:
\begin{equation}
u_\mu T^{\mu \nu}_{\hspace{10pt}; \nu}\equiv u_\mu \left[ \frac{1}{a^3}\frac{\partial }{\partial x^\nu}(a^3T^{\mu \nu})+\Gamma^{\mu}_{\alpha \beta}T^{\alpha \beta}\right],
\end{equation}
with $T^{\mu \nu} = T^{\mu \nu}_{coll}  + \Delta {\tilde T^{\mu \nu}}$, where $T^{\mu \nu}_{coll}$ is given by  Eq. (\ref{KF3}). Now,  by summing over the repeated indices and using (\ref{symbols}) it becomes: 
\begin{eqnarray}
u_\mu T^{\mu \nu}_{\hspace{10pt}; \nu}&\equiv&\frac{1}{a^3}\frac{\partial}{\partial t}(a^3T^{00}_{coll})+\Gamma^0_{ij}(T^{ij}_{coll} + \Delta {\tilde T^{ij}}) \nonumber \\
&=&\frac{1}{a^3}\frac{\partial}{\partial t}(a^3\int{f E d^3p})+3\frac{\dot a}{a}(P +{\tilde P}) \nonumber \\
&=& 3H(\rho + P + {\tilde P}) + H\left(1-\frac{\Gamma}{\Theta} \right)\int{Ep\frac{\partial f}{\partial p}d^3p} \nonumber \\
&=& 3H(\rho + P + {\tilde P}) - 3H\left(1-\frac{\Gamma}{\Theta} \right)(\rho + P) \nonumber \\
&=& 3H{\tilde P} + (\rho+P)\Gamma.
\end{eqnarray}
Therefore, we obtain a divergenceless total energy momentum-tensor, as required by the Einstein gravitational equations, only if the creation pressure ${\tilde P}$ is given by:
\begin{equation}
{\tilde P}=-(\rho + P)\frac{\Gamma}{\Theta} \equiv P_c,
\end{equation}
in complete agreement with the macroscopic expression [see Eq.(\ref{pressaotermo})]. 

An alternative simplified kinetic deduction of the above expression can be seen in the Appendix B (see also comment \cite{EMT}). 
\end{itemize}

\section{Temperature Evolution Law}       

Let us now proceed to calculate the temperature evolution for cosmic fluids endowed with ``adiabatic" gravitational particle production based on our kinetic approach. In this section we assume that the ``adiabatic" distribution function is given by the same equilibrium form (see previous section and Appendix A).  
\begin{equation}
f=e^{\alpha(t)-\beta(t) E},
\end{equation} 
where $\alpha(t)$ is a scalar function and $\beta(t)$ can be interpreted as the inverse of temperature. Inserting  this ``adiabatic" form into the Boltzmann's equation (\ref{boltzmann}) we obtain: 
\begin{equation}
 \dot \alpha -\dot \beta E+ \beta H\left( 1-\frac{\Gamma}{\Theta} \right)\frac{p^2}{E}=0.
\end{equation}
This equation has a trivial solution  given by $\dot \alpha = {\dot \beta}=0$  and  $\Gamma = \Theta = 3H$. It represents an eternal exponential de Sitter solution since from Eq. (\ref{dotrhoA}) we find $\dot \rho = {\dot n} = 0$. Physically, it means that the matter creation rate
has exactly the value that compensates for the dilution of particles due to expansion. Such a solution was macroscopically  derived long ago by Lima, Germano and Abramo \cite{LGA96}.  

Now, for $\Gamma \neq \Theta$, the above differential equation has also two extreme limiting solutions. The ultrarelativistic or negligible mass limit  ($m \ll T, E \simeq p$), and the nonrelativistic limit ($m\gg T$, $E \simeq m+\frac{p^2}{2m}$). Let us now determine the solutions for such limits. 
\begin{itemize}

\item $m \ll T$

\end{itemize}
In this case the above equation becomes:

\begin{equation}
 \frac{\dot \alpha}{\dot \beta}=E\left[1-\left( 1-\frac{\Gamma}{\Theta}\right)\frac{\dot a}{a}\frac{\beta }{\dot \beta} \right],
\end{equation}
which has the solution $\dot \alpha =0$ with the solution for $\beta = 1/T$ recovering the non-equilibrium thermodynamic result [see Eq.(28)]
\begin{equation}\label{Tempm_0}
 \frac{\dot T}{T}=-\frac{\dot a}{a}+\frac{\Gamma}{3} \Leftrightarrow \frac{1}{aT}\frac{d (aT)}{dt} = \frac{\Gamma}{3}.
\end{equation}

By fixing the constant at the present time, the general solution of the above equation can be written as:
\begin{equation}\label{temperaturethermo1}
T=T_0 \left( \frac{a_0}{a} \right) e^{{\frac{1}{3}\int^{t_o}_{t}{\Gamma(t')}{dt'}}}.
\end{equation}
In the simplest but interesting case, the creation rate $\Gamma$ remains constant for a given cosmic time interval. This kind of situation may happens at the early inflation phase or at late times of the evolution.  By defining $\Delta t= t_f - t_i$ one finds the general solution:

\begin{equation}\label{temperaturethermo}
T_f=T_i\left(\frac{a_i}{a_f}\right)e^{\frac{\Gamma}{3}(t_f - t_i)}.
\end{equation}

Now, in terms of the redshift, the general solution (\ref{temperaturethermo1}) for the CMB fluid endowed with ``adiabatic'' photon creation is given by (see also discussions below (\ref{eq:EVOLT}) and (\ref{temp2})):
\begin{equation}\label{NEQT}
T=T_0(1+z)e^{{\frac{1}{3}}{\int^{z}_0{\Gamma(z')\frac{dt}{dz'}dz'}}},
\end{equation}
and, as should be expected, for $\Gamma=0$, the same equilibrium result is recovered.

It is also useful to show how the above temperature law (\ref{Tempm_0}) for effectively massless  particles $(m \ll T)$ or radiation determines the energy density. By eliminating $\Gamma$ from equations (\ref{Tempm_0}) and (\ref{dotrhoA}) we find:
\begin{equation}\label{energyL}
 \frac{\dot \rho}{\rho}= 4\frac{\dot T}{T} \,  \Rightarrow \, \rho \propto T^{4}.
\end{equation}
The above equation was first determined based on thermodynamics, but now it has been recovered from a kinetic approach. It means that under``adiabatic" conditions particles are created but the energy density as a function of the temperature is given by the same expression of equilibrium. Indeed, by using this result one may show that the spectrum of radiation is also preserved in the course of the cosmic evolution \cite{Lima97}.

\begin {itemize} 
\item $m \gg T$

\end{itemize} 
In this limit, the equation becomes
\begin{equation}
 \frac{\dot \alpha}{\dot \beta}-m=\frac{p^2}{m}\left[ \frac{1}{2} - H \frac{\beta }{\dot \beta}\left( 1-\frac{\Gamma}{\Theta} \right) \right],
\end{equation}
and has the solution $\alpha - m \beta = \textrm{const}$ and
\begin{equation}
 \frac{\dot T}{T}=-2\left( \frac{\dot a}{a} - \frac{\Gamma}{3}\right),
\end{equation}
which is the same macroscopic law as given by (\ref{temp3}). 

As one may check, in terms of the redshift, the general solution $T(z)$ for a nonrelativistic fluid  endowed with ``adiabatic'' particle creation  reads: 
\begin{equation}\label{temperaturethermo}
T=T_0(1+z)^{2}e^{{\frac{1}{3}}{\int^{z}_0{\Gamma(z')\frac{dt}{dz'}dz'}}}.
\end{equation}


As an illustration of the developed formalism we may consider the phenomenological law ${\Gamma}=3\beta H$  often discussed in the literature \cite{Models,TempL}. In this case it is easy to see from (53) and (57), as well as from the general solutions in terms of the redshift  parameter (see Eqs. (55) and (58)) that the  temperature relations become: 
\begin{equation}
 T=T_0(1+z)^{1 - \beta} \,\,\,for\,\, m=0,
\end{equation}
while for non-relativistic  particles the result is
\begin{equation}
T=T_0 (1+z)^{2(1 - \beta)},\,\,\,for\,\, m >> T.
\end{equation}
Naturally, when $\beta=0$ (no production), the usual results are recovered.

\section{Concluding Remarks}

In this paper we have investigated the thermodynamic and kinetic 
properties of the irreversible gravitationally induced particle production process in 
the framework of homogeneous and isotropic FRW geometries.

We showed that the macroscopic process named  ``adiabatic'' particle production 
has a consistent kinetic counterpart which can also be termed``adiabatic" in the sense that the entropy increases but the specific entropy remains constant \cite{LCW}.  

In this way, all the results obtained in the macroscopic approach were recovered by the new kinetic treatment proposed here. The temperature law was determined for two important limits, namely, the ultrarelativistic ($m << T$) and nonrelativistic ($m >> T$) domains.  It was shown that the energy density of massless or ultrarelativistic particles is also proportional to $T^{4}$ as demonstrated long ago based  on the macroscopic approach. In particular, this means that the early proof showing that the CMB spectrum is not destroyed when the cosmic creation process happens under ``adiabatic'' conditions \cite{Lima97} has now been confirmed based on the relativistic Boltzmann equation including  particle production as modeled here. In this connection,  it is worth mentioning that the gravitationally induced production of massless particles is forbidden during the radiation phase of  FRW models  described in the context of general relativity. However, this is not true for alternative theories, as recently discussed in the context of F(R) cosmologies \cite{F(R)}, and, as such, it should also be interesting to investigate the kinetic creation process in a more general gravitational framework.    

It should be stressed that the creation kinetic treatment has been discussed here only to the case of a flat geometry. The unified general case ($k=0, \pm 1$), including a multicomponent fluid,  will be published elsewhere. In principle, by adopting a reasonable ansatz on the creation rate $\Gamma$,  the extension of the Boltzmann equation proposed here can be implemented by computational codes for physical cosmology (like CMBFAST) in order to discuss  the anisotropies of CMB  in the presence of gravitationally induced particle production. In this connection, we notice that the kinetic counterpart of the so-called  creation cold dark matter (CCDM) cosmology proposed earlier based on the macroscopic approach was presented in the Appendix D. 

\vspace{0.3cm}

\appendix 
\section {On the Divergence of the Entropy Flux}

In  section $\ref{kinetic}$ no restriction to the distribution function were done to show that the  divergence of the entropy flux is $S^{\mu}_{; \mu}=s \Gamma$. Now, we will show that when  the ``adiabatic" form  of the distribution function is adopted  the same result is obtained. By inserting $f=e^{\frac{\mu}{T}-\frac{E}{T}}$ into (\ref{entropyflux}), doing the identifications, $\alpha=\frac{\mu}{T}$ and $\beta=\frac{1}{T}$,  and calculating the divergence it follows that:
\begin{eqnarray}
 S^{\mu}_{;\mu}=-3H\int{(f \ln f - f)d^3p}-\int{\frac{\partial f}{\partial t}\ln f d^3p}.
\end{eqnarray}
Hence, by noting that $s=-\int{(f \ln f - f)d^3p}$ and using (\ref{boltzmann}), we obtain
\begin{eqnarray}
 S^{\mu}_{;\mu}= \left[\frac{\rho + P - n\mu}{T}\right]\Gamma = s\Gamma,
\end{eqnarray}
as should be expected. Note that in the the last step we have used the local form of Euler's relation (\ref{entropyd}). 

\vspace{1.0cm}
\section {On the non-collisional creation pressure}

Here we show the simplest manner to obtain  kinetically the creation pressure. Firstly, we multiply  by E the extended Boltzmann equation (\ref{boltzmann}) and integrate it over the  momentum space to obtain:  

\begin{equation}\label{DIV1}
\int{E\frac{\partial f}{\partial t}d^3p}-H\left(1-\frac{\Gamma}{\Theta} \right)\int{Ep\frac{\partial f}{\partial p}d^3p} = 0 
\end{equation}
Now, by integrating the above equation term by term we find: 

\begin{equation}\label{DIV}
{\dot \rho} + \Theta(\rho + P) -\Gamma (\rho + P) = 0.
\end{equation}
or equivalently,
\begin{equation}\label{DIV2}
{\dot \rho} + \Theta(\rho + P + P_c) = 0 \Leftrightarrow P_c = - (\rho + P)\frac{\Gamma}{\Theta}. 
\end{equation}
which is the energy conservation including the definition of the creation pressure as given by (\ref{energy}) and (\ref{pressaotermo}). 

It is also interesting that integrating the extended Boltzmann equation (\ref{boltzmann}) over $d^{3}p$ and combining it with the result (\ref{DIV}) one obtains: 
\begin{equation}\label{dotrhoA}
\frac{\dot {\rho}}{\rho + P} = \frac{\dot n}{n} \equiv \Gamma - \Theta,
\end{equation}
which reproduces the macroscopic constraint given by (\ref{dotrho}).

\section{Kinetic CCDM Cosmology}

As remarked in the introduction, the alternative scenario mimicking $\Lambda$CDM model usually referred to as creation cold dark matter (CCDM) cosmology \cite{LJO2010},  was proposed based on the macroscopic irreversible covariant approach discussed in section IIB.  Now we show that it can also be formulated based on a kinetic theoretical framework. First we assume that baryons are conserved and that the ratio $\Gamma/\Theta$ takes the form:

\begin{equation}
\label{Gamma} 
\frac{\Gamma}{3H}=\alpha \frac{\rho_{co}}{\rho_{dm}},
\end{equation}
where $\alpha$ is a constant parameter, ${\rho_{co}}$ is the present day value of the critical density.

Now, combining Friedman equation
\begin{equation}
8\pi G(\rho_{dm} + \rho_b) = 3H^{2},
\end{equation} 
with the energy conservation law for both components one finds:

\begin{equation}
\label{HzFlat}
 \left(\frac{H}{H_0}\right)^2= \Omega_{meff}(1+z)^3+\alpha,
\end{equation}
where the effective nonrelativistic matter density parameter, $\Omega_{meff}= \Omega_{dm} + \Omega_{b} - \alpha \equiv 1 - \alpha$,  describes the clustered matter. Now, comparing with the cosmic concordance prediction:

\begin{equation}
\label{HzFlat1}
 \left(\frac{H}{H_0}\right)^2= \Omega_m(1+z)^3+ \Omega_{\Lambda},
\end{equation}
we see that the creation dimensionless parameter in (\ref{HzFlat}) plays the role of an effective  density  parameter ($\alpha \equiv \tilde \Omega_{\Lambda}$) associated with the effective $\Lambda$-term ($\tilde \Lambda \equiv 3\alpha H_0^{2}$). Like in the in the  standard flat  $\Lambda$CDM model, it should be noticed that $\alpha$ is the only dynamic free parameter (for more details see Refs. \cite{LJO2010} and \cite{Waga2014}).   

\vspace{0.2cm}

\begin{acknowledgments}
The authors are grateful to Gary Steigman for stimulating discussions on kinetic theory. JASL and IB are partially supported by CNPq and FAPESP (Brazilian
Research Agencies).
\end{acknowledgments}


\begin{thebibliography}{99}
\bibitem{EMT} In the presence of a non-collisional external source, the collisional kinetic EMT alone is not conserved. Observe that we have obtained a divergenceless EMT  working in the same spirit of classical electrodynamics when only the total energy momentum tensor including field and particles is conserved. It is easy to show that in our approach $u_\mu {T^{\mu \nu}_{coll}}_{;\nu} = \Gamma (\rho + p)$. Therefore, one may also prefer simply to define an equivalent EMT identically conserved by moving the term $\Gamma (\rho + P)$ to the RHS of the energy balance equation. Finally, the negative contribution must be identified with $P_c\Theta$ from macroscopic theory (see Appendix B for a different approach).
\end{thebibliography}
\end{document}